\documentclass{PoS}

\newcommand{\MeV}{\textrm{MeV}}

\usepackage{amsmath}
\usepackage{enumerate}
\usepackage{float}
\usepackage[utf8]{inputenc}

\title{Parametrized Equation of State for QCD from 3D Ising Model}

\ShortTitle{Parametrized Equation of State for QCD from 3D Ising Model}

\author{\speaker{Paolo Parotto}%
       
       \\
       Department of Physics, University of Houston, Houston TX, USA\\
       E-mail: \email{pparotto@uh.edu}}

\author{}

\abstract{The only first principle knowledge of the QCD equation of state at finite baryonic density is given from Lattice QCD as a Taylor expansion around $\mu_B = 0$. The coefficients of such an expansion are currently available up to order ${\cal O}(\mu_B^6)$. The expected critical behavior of QCD is in the same static universality class as the 3D Ising model. By means of a suitable parametrization for the scaling equation of state of 3D Ising and a parametrized map to connect to QCD, we present an equation of state matching first principle Lattice QCD calculations, which spans the values of baryonic densities explored in the BES-II program, and includes the correct scaling behavior in the proximity of the critical point. 

This EoS can serve as an important ingredient for the fluid dynamical simulations of heavy ion collisions at BES energies needed as a basis for the calculation of observables. Future comparisons between such calculations and BES-II data can constrain the parameters in the EoS -- including the parameters that describe the location of the critical point.

This contribution reports on work done within the Fluctuations/Equation of State working group of the BEST Collaboration.}

\FullConference{Critical Point and Onset of Deconfinement\\
                 7-11 August, 2017\\
                 The Wang Center, Stony Brook University, Stony Brook, NY}

\begin{document}

\section{Introduction}
Determining thermodynamic properties of strongly interacting matter is a crucial goal of high energy nuclear physics, and strong efforts are currently in place from both theory and experiment to expand the knowledge of the different phases of QCD matter. The reach of such knowledge would range from giving a better understanding of the first microseconds of evolution of the universe, to represent a major step in the study of cold and dense astrophysical objects such as neutron stars. 

Over the past decades, countless models have been produced in order to study and describe the behavior of strong matter in its different phases and regimes. In the last 15 years, first principle Lattice calculations have given precise quantitative results in the baryon-antibaryon symmetric regime -- the one of the early universe -- over a broad range of temperatures, confirming the presence of a continuous phase transition at vanishing baryon density at a temperature of $T \simeq 155 \, \MeV$ \cite{Aoki:2006we, Aoki:2009sc, Borsanyi:2010bp, Bazavov:2011nk}.

It is now strongly believed \cite{Asakawa:1989bq,Halasz:1998qr,Berges:1998rc,Stephanov:2004wx,Stephanov:2007fk} that at some higher baryonic densities, the chiral/deconfinement transition would be of the first order, thus implying the presence of a critical point. Past studies \cite{Halasz:1998qr,Berges:1998rc,Pisarski:1983ms} have shown that such a critical point would be in the same universality class as the three dimensional Ising model. The experimental search of this critical point, as well as the continuous theoretical effort to interpret experimental results in this search, have currently reached a peak of productivity in light of the BES-II program, which will take place at the Relativistic Heavy Ion Collider in the next couple of years, exploring the higher density region of the QCD phase diagram. The main intent of the program is to locate the critical point, or alternatively rule out its presence in a broad range of densities.

Hydrodynamic simulations play a major role in the study of heavy-ion collisions, and therefore the theoretical interpretation of experimental data. The main ingredient needed in order to perform a hydrodynamic simulation is an equation of state of QCD matter that would drive the evolution of the system. Therefore, in studying the potential presence of a critical point in the range of densities accessible to the BES-II program, the need of an equation of state which includes critical behavior is indisputable. Whereas models have been used to produce such an equation of state, the purpose of this work is to generate an equation of state in a parametric form, which contains critical behavior \textit{in the right universality class}, and on the other hand \textit{matches the known first principle Lattice QCD results at vanishing baryon density}. 

Current knowledge of QCD equation of state (EoS) from first principle calculations is in the form of a Taylor expansion of the pressure in the baryonic chemical potential around $\mu_B=0$. This is due to the well-known sign problem of Lattice simulations of QCD at finite density. Hence, the EoS can be given as \cite{Bazavov:2017dus,Bellwied:2015rza}:
\begin{equation}
\frac{P \left( T, \mu_B \right)}{T^4} = \sum_n c_{2n} (T) \left( \frac{\mu_B}{T} \right)^{2n} \, \, , \qquad  \qquad c_n (T) = \left. \frac{1}{n!} \frac{\partial^n P/T^4}{\partial (\mu_B/T)^n} \right|_{\mu_B=0}
\end{equation} 

There have been studies \cite{Gavai:2004sd,Karsch:2011yq,DElia:2016jqh,Bazavov:2017dus} aimed at determining whether, and under what requirements, it would be possible to extract information about a possible critical point just from Lattice QCD data, however the number of coefficients in the expansion currently allows to only partially constrain the location of such point. Moreover, the extent of validity of the Taylor series can never reach beyond the baryonic chemical potential at the critical point. On the other hand, the fact that the universality class to which critical point belongs is known, allows one to impose the behavior of the EoS in some region of a certain size around the critical point. \\

The strategy pursued in this work can be summarized as follows:
\begin{enumerate}[i)]
\item Make use of a suitable parametrization to describe the universal scaling behavior of the EoS in the 3D Ising model near the critical point;
\item Map the 3D Ising model phase diagram onto the one of QCD via a parametric, non universal change of variables;
\item Use the thermodynamics of the Ising model EoS to estimate the critical contribution to the expansion coefficients from Lattice QCD; 
\item Reconstruct the full pressure, matching Lattice QCD at $\mu_B=0$ and including the correct critical behavior.
\end{enumerate} 

Note that the parametric nature of an EoS constructed with the described strategy has the advantage of allowing the influence of the presence of a critical point on the thermodynamics itself and on hydrodynamic simulations, as well as the disadvantage of relying on a number of parameters. However, the choice of parameters in the Ising $\longmapsto$ QCD map is not free. 

Current knowledge from Lattice QCD results already puts constraints on the location of the critical point as well as other parameters (more details will follow in the next section). Moreover, thermodynamic consistency requirements will have to be met by the produced EoS, thus reducing the possible choice of parameters allowed.

\section{Scaling EoS in 3D Ising model and map to QCD}
For the scaling EoS of the 3D Ising model, in a neighborhood of the critical point, one can use the following parametrization for the magnetization $M$, the magnetic field $h$ and the reduced temperature $r=\left( T- T_C \right)/T_C$ \cite{Nonaka:2004pg,Guida:1996ep,Schofield:1969zz,Bluhm:2006av}:
\begin{align}
M &= M_0 R^\beta \theta \, \, , \\
h &= h_0 R^{\beta \delta} \tilde{h}(\theta) \, \, ,  \\
r &= R (1- \theta^2) \, \, .
\end{align}
where $M_0$, $h_0$ are normalization constants, $\tilde{h}(\theta) = \theta (1 + a \theta^2 + b \theta^4)$  with $a=-0.76201$, $b=0.00804$, $\beta$ and $\delta$ are 3D Ising critical exponents \cite{Guida:1996ep}, and the parameters take on the values $R \geq 0$, $\left| \theta \right| \leq \theta_0 \simeq 1.154$, $\theta_0$ being the first non-trivial zero of $\tilde{h}(\theta)$.

\begin{figure}[h]
\includegraphics[width=\textwidth]{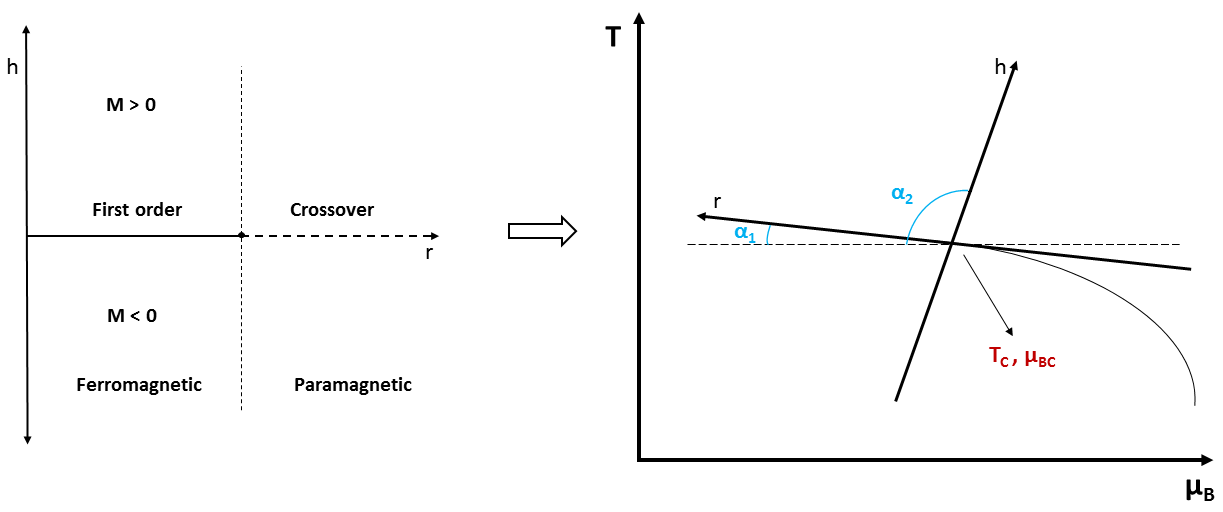}
\caption{The 3D Ising model phase diagram is mapped onto the QCD one by means of a linear transformation.}
\label{fig:IsQCD}
\end{figure}

Following \cite{Nonaka:2004pg, Guida:1996ep}, one can write the Gibbs free energy density as a function of the parameters $(R,\theta)$ and, using the thermodynamic relation $G/V = g = - P$ between Gibbs free energy density and pressure, write down the latter in the Ising model scaling EoS as:
\begin{equation}
P_{\text{Ising}}(R, \theta) = - h_0 M_0 R^{2 - \alpha} \left[ g(\theta) - \theta \tilde{h}(\theta) \right] \, \, .
\end{equation} 

In order to transfer the critical thermodynamics to QCD, a non-universal mapping is needed between Ising variables $(h,r)$ and QCD coordinates $(T,\mu_B)$  (see Fig. \ref{fig:IsQCD}). The most general linear transformation allowing this makes use of six parameters:
\begin{align}
\frac{T - T_C}{T_C} &=  w \left( r \rho \,  \sin \alpha_1  + h \, \sin \alpha_2 \right) \, \, , \\
\frac{\mu_B - \mu_{BC}}{T_C} &=  w \left( - r \rho \, \cos \alpha_1 - h \, \cos \alpha_2 \right) \, \, .
\end{align} 
where $(T_C,\mu_{BC})$ give the location of the critical point, $\alpha_1$ and $\alpha_2$ indicate the relative angle between the $r$ and $h$ axes and the lines of $T= \textit{const.}$, and the parameters $w$ and $\rho$ correspond to global and relative rescaling of $r$ and $h$. \\

Thanks to this transformation, it is possible to have the following map:
\begin{equation}
\left( R, \theta \right) \longmapsto \left( h, r \right) \longleftrightarrow \left( T, \mu_B \right)
\end{equation}
where the second step is globally invertible. The critical contribution to the pressure in QCD can then simply be built from:
\begin{equation}
P^{\text{QCD}}_{\text{crit}}(T, \mu_B) = f(T, \mu_B) P_{\text{Ising}} (R(T, \mu_B) ,\theta (T, \mu_B))
\end{equation} 
for some regular function $f(T, \mu_B)$ with energy dimension four. \\

Assuming it is possible to separate the critical contribution coming from the high temperature critical point in the Taylor coefficients calculated from Lattice QCD, one can write:
\begin{equation} \label{eq:coeff}
c_n^{\text{LAT}}(T) = c_n^{\text{reg}}(T) + c_n^{\text{crit}}(T) \, \, ,
\end{equation}
where on the left hand side are the coefficients calculated from Lattice QCD, hence enforcing agreement with Lattice EoS at $\mu_B=0$, and this equation should be read as a definition for the ``regular'' coefficients $c^{\text{reg}}(T)$, which will include all contributions not coming from the Ising critical point. Thus, one can reconstruct the full pressure as:
\begin{equation} \label{eq:Pfull}
P (T, \mu_B) = T^4 \sum_n c^{\text{reg}}_{2n} (T) \left( \frac{\mu_B}{T} \right)^{2n} + P^{\text{QCD}}_{\text{crit}}(T, \mu_B) \, \, . 
\end{equation}
which will, by construction, match Lattice results at $\mu_B=0$ and contain critical behavior in the correct universality class.

\section{Results}
\subsection{The choice of parameters} \label{sec:param}
Although the most general linear map between Ising variables and QCD coordinates requires the use of six parameters, it is possible to introduce some constraint in the choice by making use of additional arguments for the location of the critical point. For example, there have been works \cite{Bellwied:2015rza,Cea:2015bxa,Bonati:2015bha,Bazavov:2017dus} that have calculated the curvature of the crossover line of the chiral transition at $\mu_B=0$, approximating the shape of such transition line with a parabola:
\begin{equation}
T = T_0 + \kappa \, T_0 \left(\frac{\mu_B}{T_0}\right)^2 + {\cal O} (\mu_B^4)
\end{equation}
where $T_0 \simeq 155 \, \text{MeV}$ and $\kappa \simeq -0.0149$ (the values are from \cite{Bellwied:2015rza}) are the transition temperature and curvature of the transition line at $\mu_B=0$, respectively. The number of the parameters is thus reduced to 4, being the angle $\alpha_1$ also fixed by:
\begin{equation}
\alpha_1 = \tan^{-1} \left( 2 \frac{\kappa}{T_0} \mu_{BC} \right) \, \, .
\end{equation} 

The aim of the EoS being to be employed in hydrodynamic simulations for heavy-ion collisions in the BES-II program, we will make a choice to have a value of the baryonic chemical potential which is accessible within such program, hence in the following we will set $\mu_{BC} = 350 \, \MeV$, resulting in:
\begin{align} 
T_C \simeq 143.2 \, \MeV \, , \qquad \qquad \alpha_1 \simeq 3.85 \, ^\circ \, \, .
\end{align}
In addition, the axes are chosen to be orthogonal, so that $\alpha_2 \simeq 93.85 \, ^\circ$, and the scaling parameters are:
\begin{align}
w = 1 \, \, , & \qquad \qquad \rho = 2 \, \, .
\end{align}

\subsection{Parametrization of Lattice data}
For the purpose of being used in hydrodynamic simulation, our EoS needs to cover the region of the phase diagram at low temperature, which is not available lattice simulations (typically $T \simeq 100 \, \MeV$ is the lower bound for Lattice results). In order to solve this issue, we extend the Lattice data downwards in temperature by calculating the baryon susceptibilities that appear as coefficients in the Taylor expansion making use of the ideal Hadron Resonance Gas (HRG) model, which is commonly accepted as a good approximation of QCD in this regime. 

\begin{figure}[h]
\center
\includegraphics[width=.32\textwidth]{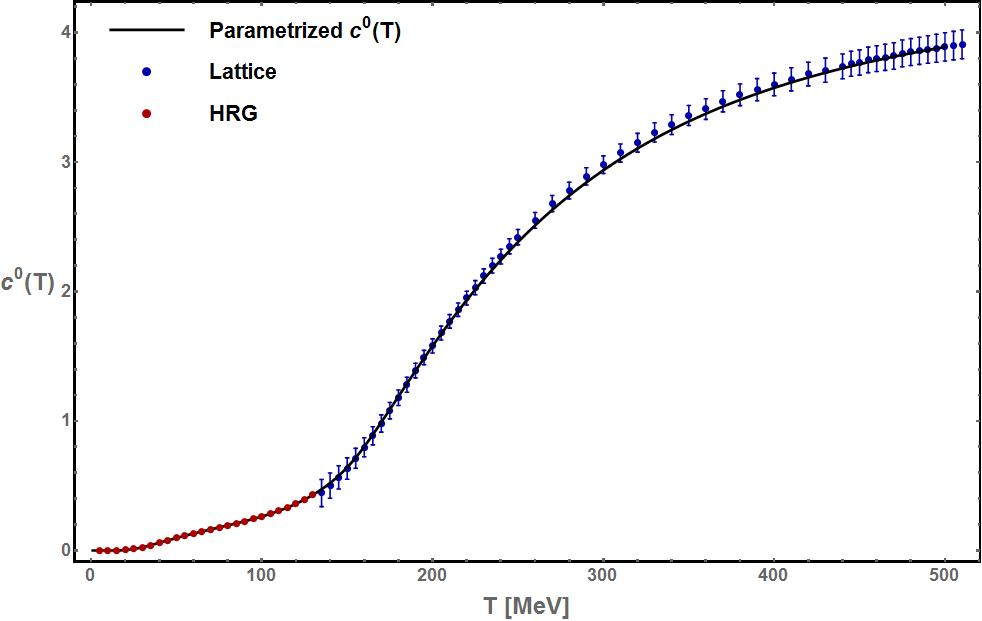}
\includegraphics[width=.32\textwidth]{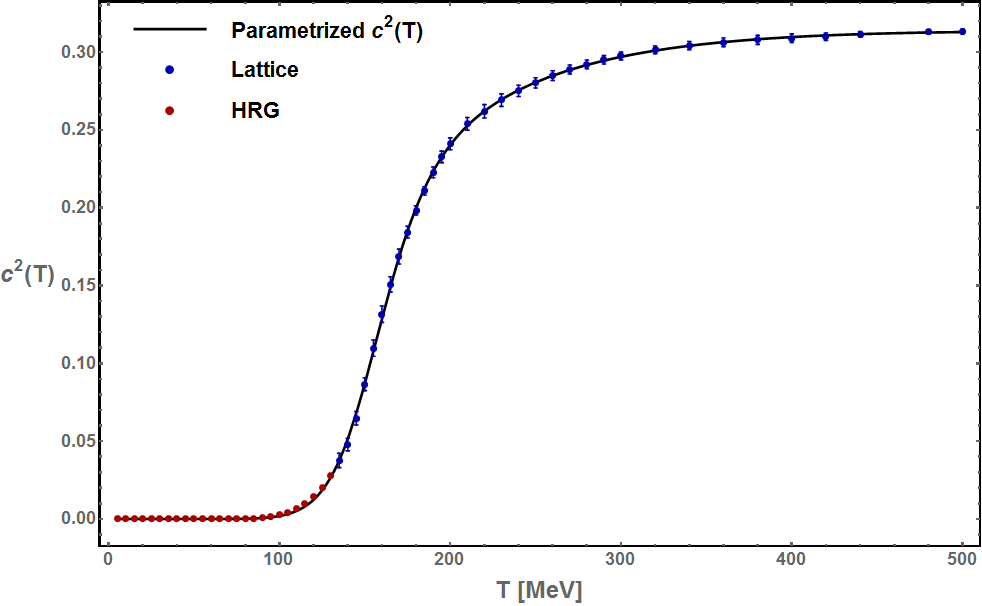} 
\includegraphics[width=.32\textwidth]{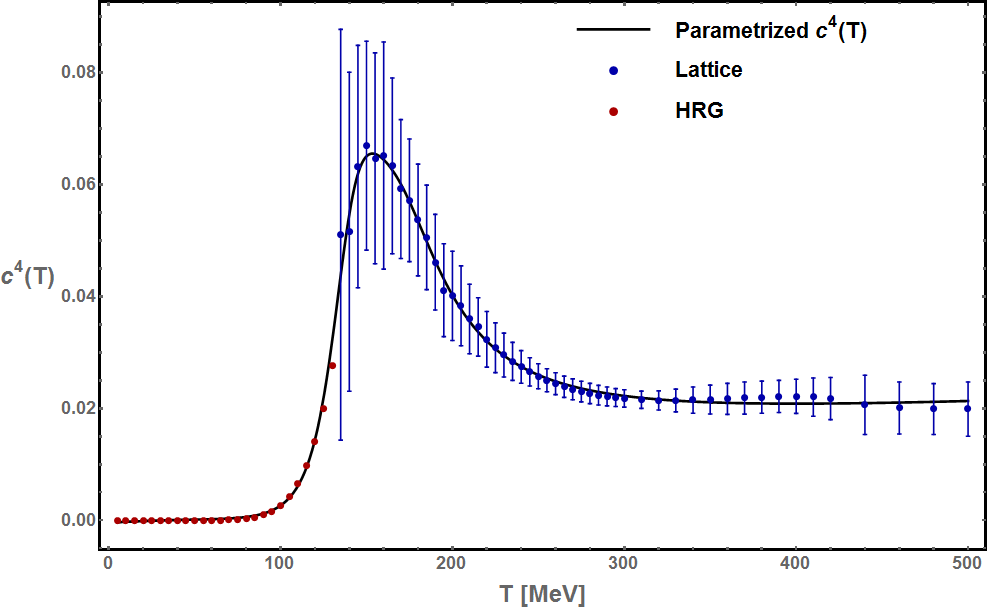}
\caption{Parametrization of baryon susceptibilities from Lattice QCD \cite{Bellwied:2015rza} and HRG model calculations.}
\label{fig:param}
\end{figure}

In addition, the resulting data from Lattice/HRG are parametrized in order to obtain a dependence on the temperature which is  smooth enough to obtain tractable results for the entropy density and baryon density, which are first derivatives of the pressure. The parametrization is performed in the range $T = 5 - 500 \, \MeV$ via a ratio of $5^{\text{th}}$ order polynomials in the inverse temperature (see Fig.\ref{fig:param}).

\begin{figure}[h]
\center
\includegraphics[width=.32\textwidth]{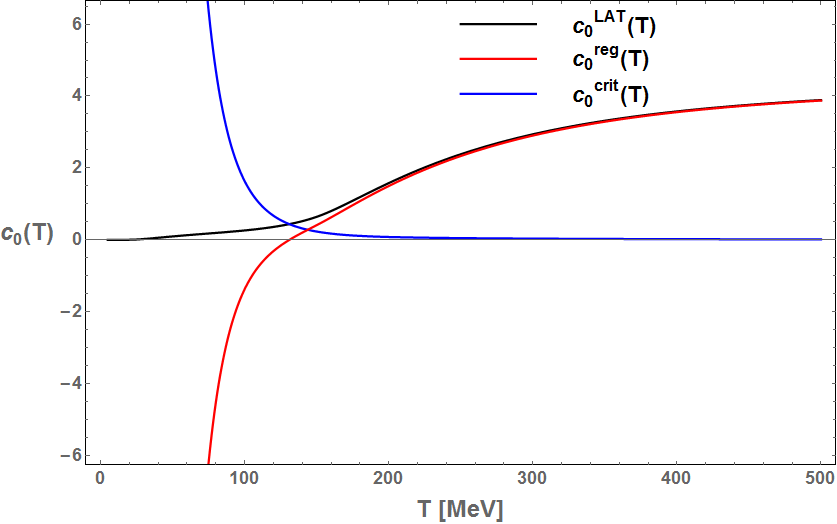}
\includegraphics[width=.32\textwidth]{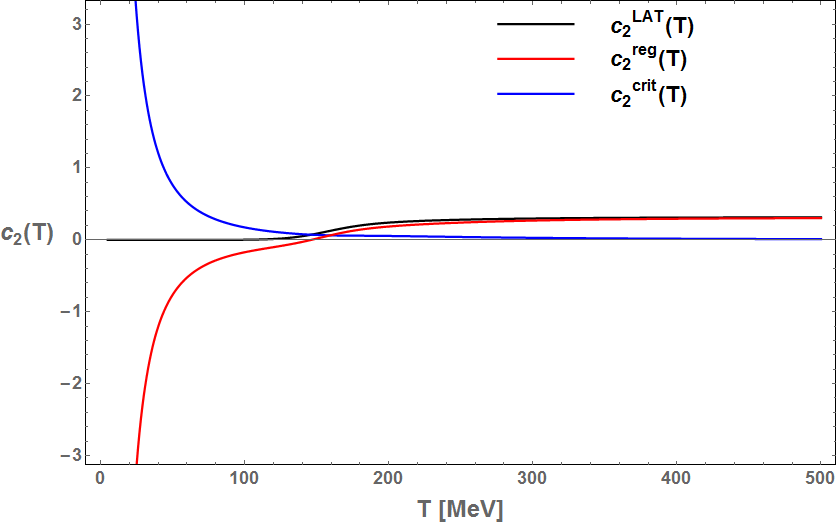} 
\includegraphics[width=.32\textwidth]{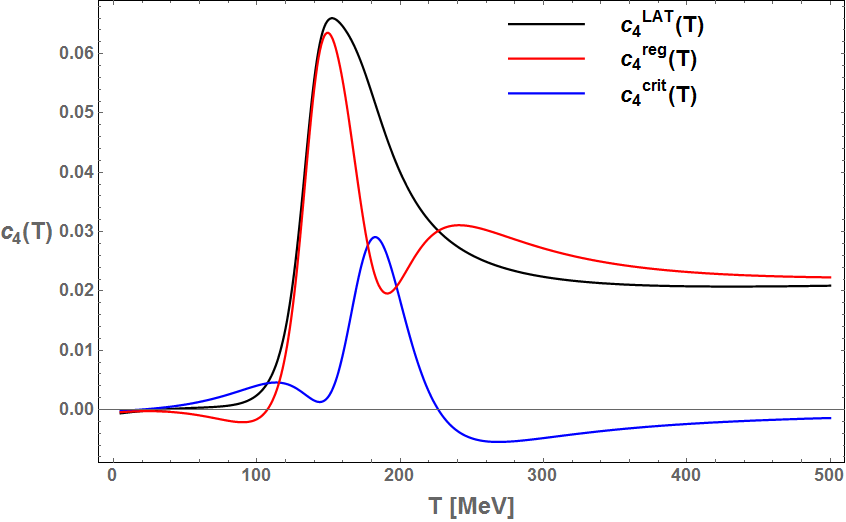} \\
\caption{Comparison of critical (blue) and non-critical (red) contributions to baryon susceptibilities up to ${\cal O}(\mu_B^4)$ with parametrized Lattice data from the Wuppertal-Budapest collaboration \cite{Bellwied:2015rza}.}
\label{fig:chis}
\end{figure}

The smooth curves obtained with the parametrization will be the $c_n^{\text{LAT}} (T)$ coefficients in Eq. (\ref{eq:coeff}), thus defining the $c_n^{\text{reg}} (T)$ coefficients that will be used for the Taylor expansion. In Fig.\ref{fig:chis} we can see the comparison of the critical and ``regular'' contribution with the parametrized Lattice/HRG model results.

\begin{figure}[h]
\center
\includegraphics[width=.8\textwidth]{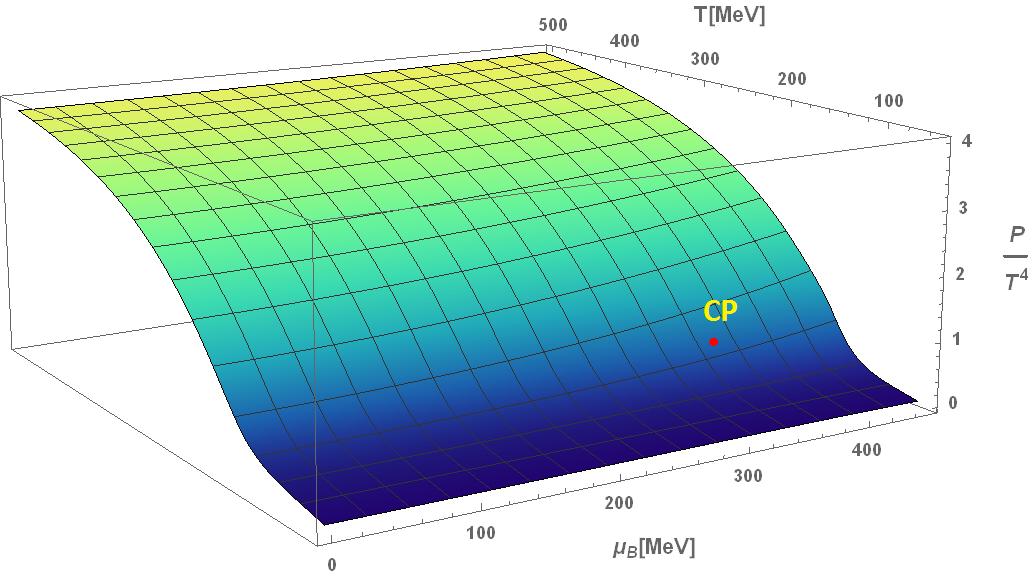}
\caption{Full pressure for the choice of parameters in Section \ref{sec:param}}
\label{fig:Pfull}
\end{figure}

\begin{figure}[h]
\center
\includegraphics[width=.8\textwidth]{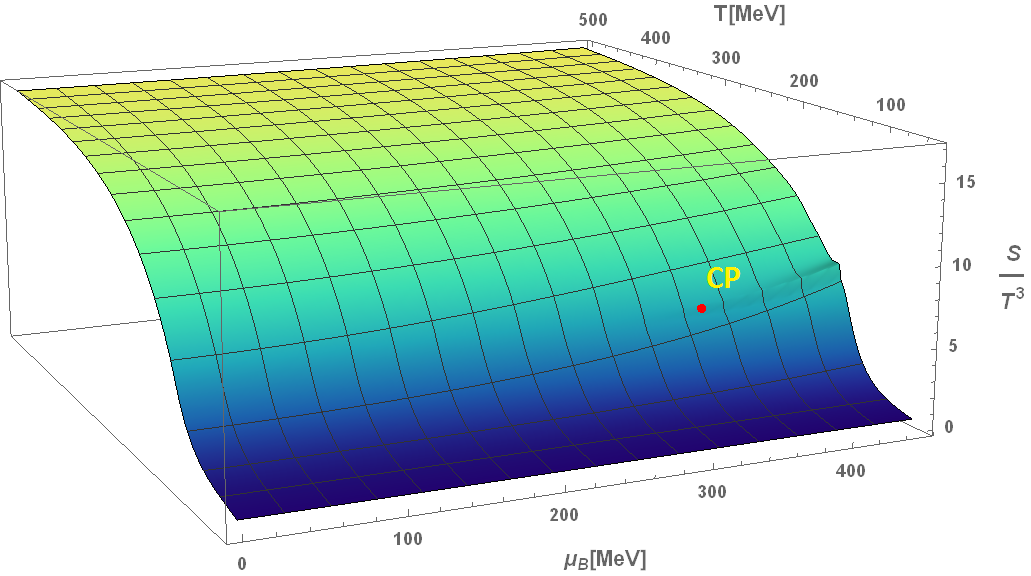}
\caption{Entropy density for the choice of parameters in Section \ref{sec:param}}
\label{fig:Entrfull}
\end{figure}

The reconstruction of the full pressure is now straightforward, and can be carried out as in Eq.(\ref{eq:Pfull}):
\begin{equation}
P (T, \mu_B) = T^4 \sum^2_{n=0} c^{\text{reg}}_{2n} (T) \left( \frac{\mu_B}{T} \right)^{2n} + T_C^4 \, P^{\text{QCD}}_{\text{Ising}}(T, \mu_B) \, \, ,
\end{equation}
which is shown, with the current choice of parameters and up to order ${\cal O} (\mu_B^4)$, in Fig.\ref{fig:Pfull}, for $T = 50 - 500 \, \MeV$ and $\mu_B = 0 - 450 \, \MeV$. The entropy density, defined from the pressure as:
\begin{equation}
S(T,\mu_B) = \left( \frac{\partial P(T,\mu_B)}{\partial T} \right)_{\mu_B} \, \, ,
\end{equation}
is shown in Fig. \ref{fig:Entrfull}, where the discontinuity due to the first order phase transition for $\mu_B > \mu_{BC}$ is visible.
 
\section{Discussion}
A parametrized equation of state for QCD that matches Lattice results \textit{exactly} at vanishing baryochemical potential and contains critical behavior in the expected universality class of the theory is presented in this work.

By means of a parametrization for the scaling EoS in the vicinity of the Ising-like critical point and of a non-universal map from the Ising model variables to QCD coordinates, it was possible to calculate the critical contribution to thermodynamic quantities (i.e. the pressure) in QCD at $\mu_B=0$. The reconstructed pressure in Fig.\ref{fig:Pfull}, along with the entropy density in Fig. \ref{fig:Entrfull}, can be readily used, together with other thermodynamic quantities (baryon density, energy density, speed of sound, etc.), in hydrodynamic simulations of heavy ion-collisions.  

When experimental data from the BES-II program will be available, the comparison of such data with predictions (e.g. baryon number fluctuation observables) from hydrodynamical simulations that make use of the presented EoS, will constrain the values of the parameters employed, thus possibly provide indication on the location of the critical point.

\section*{Acknowledgements}
This material is based upon work supported by the National Science Foundation under Grants No. PHY-1513864, PHY-1654219 and OAC-1531814 and by the U.S. Department of Energy, Office of Science, Office of Nuclear Physics, within the framework of the Beam Energy Scan Theory (BEST) Topical Collaboration. An award of computer time was provided by the INCITE program. This research used resources of the Argonne Leadership Computing Facility, which is a DOE Office of Science User Facility supported under Contract No. DE-AC02-06CH11357. The authors gratefully acknowledge the use of the Maxwell Cluster and the advanced support from the Center of Advanced Computing and Data Systems at the University of Houston.

\bibliography{CPOD2017_proceeding}

\providecommand{\href}[2]{#2}\begingroup\raggedright\begin{thebibliography}{10}

\bibitem{Aoki:2006we}
Y.~Aoki, G.~Endrodi, Z.~Fodor, S.~D. Katz and K.~K. Szabo, \emph{{The Order of
  the quantum chromodynamics transition predicted by the standard model of
  particle physics}}, \href{https://doi.org/10.1038/nature05120}{\emph{Nature}
  {\bfseries 443} (2006) 675--678},
  [\href{https://arxiv.org/abs/hep-lat/0611014}{{\ttfamily hep-lat/0611014}}].

\bibitem{Aoki:2009sc}
Y.~Aoki, S.~Borsanyi, S.~Durr, Z.~Fodor, S.~D. Katz, S.~Krieg et~al.,
  \emph{{The QCD transition temperature: results with physical masses in the
  continuum limit II.}},
  \href{https://doi.org/10.1088/1126-6708/2009/06/088}{\emph{JHEP} {\bfseries
  06} (2009) 088}, [\href{https://arxiv.org/abs/0903.4155}{{\ttfamily
  0903.4155}}].

\bibitem{Borsanyi:2010bp}
{\scshape Wuppertal-Budapest} collaboration, S.~Borsanyi, Z.~Fodor,
  C.~Hoelbling, S.~D. Katz, S.~Krieg, C.~Ratti et~al., \emph{{Is there still
  any $T_c$ mystery in lattice QCD? Results with physical masses in the
  continuum limit III}},
  \href{https://doi.org/10.1007/JHEP09(2010)073}{\emph{JHEP} {\bfseries 09}
  (2010) 073}, [\href{https://arxiv.org/abs/1005.3508}{{\ttfamily 1005.3508}}].

\bibitem{Bazavov:2011nk}
A.~Bazavov et~al., \emph{{The chiral and deconfinement aspects of the QCD
  transition}}, \href{https://doi.org/10.1103/PhysRevD.85.054503}{\emph{Phys.
  Rev.} {\bfseries D85} (2012) 054503},
  [\href{https://arxiv.org/abs/1111.1710}{{\ttfamily 1111.1710}}].

\bibitem{Asakawa:1989bq}
M.~Asakawa and K.~Yazaki, \emph{{Chiral Restoration at Finite Density and
  Temperature}},
  \href{https://doi.org/10.1016/0375-9474(89)90002-X}{\emph{Nucl. Phys.}
  {\bfseries A504} (1989) 668--684}.

\bibitem{Halasz:1998qr}
A.~M. Halasz, A.~D. Jackson, R.~E. Shrock, M.~A. Stephanov and J.~J.~M.
  Verbaarschot, \emph{{On the phase diagram of QCD}},
  \href{https://doi.org/10.1103/PhysRevD.58.096007}{\emph{Phys. Rev.}
  {\bfseries D58} (1998) 096007},
  [\href{https://arxiv.org/abs/hep-ph/9804290}{{\ttfamily hep-ph/9804290}}].

\bibitem{Berges:1998rc}
J.~Berges and K.~Rajagopal, \emph{{Color superconductivity and chiral symmetry
  restoration at nonzero baryon density and temperature}},
  \href{https://doi.org/10.1016/S0550-3213(98)00620-8}{\emph{Nucl. Phys.}
  {\bfseries B538} (1999) 215--232},
  [\href{https://arxiv.org/abs/hep-ph/9804233}{{\ttfamily hep-ph/9804233}}].

\bibitem{Stephanov:2004wx}
M.~A. Stephanov, \emph{{QCD phase diagram and the critical point}},
  \href{https://doi.org/10.1142/S0217751X05027965}{\emph{Prog. Theor. Phys.
  Suppl.} {\bfseries 153} (2004) 139--156},
  [\href{https://arxiv.org/abs/hep-ph/0402115}{{\ttfamily hep-ph/0402115}}].

\bibitem{Stephanov:2007fk}
M.~A. Stephanov, \emph{{QCD phase diagram: An Overview}}, {\emph{PoS}
  {\bfseries LAT2006} (2006) 024},
  [\href{https://arxiv.org/abs/hep-lat/0701002}{{\ttfamily hep-lat/0701002}}].

\bibitem{Pisarski:1983ms}
R.~D. Pisarski and F.~Wilczek, \emph{{Remarks on the Chiral Phase Transition in
  Chromodynamics}}, \href{https://doi.org/10.1103/PhysRevD.29.338}{\emph{Phys.
  Rev.} {\bfseries D29} (1984) 338--341}.

\bibitem{Bazavov:2017dus}
A.~Bazavov et~al., \emph{{The QCD Equation of State to $\mathcal{O}(\mu_B^6)$
  from Lattice QCD}},
  \href{https://doi.org/10.1103/PhysRevD.95.054504}{\emph{Phys. Rev.}
  {\bfseries D95} (2017) 054504},
  [\href{https://arxiv.org/abs/1701.04325}{{\ttfamily 1701.04325}}].

\bibitem{Bellwied:2015rza}
R.~Bellwied, S.~Borsanyi, Z.~Fodor, J.~Günther, S.~D. Katz, C.~Ratti et~al.,
  \emph{{The QCD phase diagram from analytic continuation}},
  \href{https://doi.org/10.1016/j.physletb.2015.11.011}{\emph{Phys. Lett.}
  {\bfseries B751} (2015) 559--564},
  [\href{https://arxiv.org/abs/1507.07510}{{\ttfamily 1507.07510}}].

\bibitem{Gavai:2004sd}
R.~V. Gavai and S.~Gupta, \emph{{The Critical end point of QCD}},
  \href{https://doi.org/10.1103/PhysRevD.71.114014}{\emph{Phys. Rev.}
  {\bfseries D71} (2005) 114014},
  [\href{https://arxiv.org/abs/hep-lat/0412035}{{\ttfamily hep-lat/0412035}}].

\bibitem{Karsch:2011yq}
F.~Karsch, B.-J. Schaefer, M.~Wagner and J.~Wambach, \emph{{Towards finite
  density QCD with Taylor expansions}}, {\emph{PoS} {\bfseries LATTICE2011}
  (2011) 219}, [\href{https://arxiv.org/abs/1110.6038}{{\ttfamily 1110.6038}}].

\bibitem{DElia:2016jqh}
M.~D'Elia, G.~Gagliardi and F.~Sanfilippo, \emph{{Higher order quark number
  fluctuations via imaginary chemical potentials in $N_f=2+1$ QCD}},
  \href{https://doi.org/10.1103/PhysRevD.95.094503}{\emph{Phys. Rev.}
  {\bfseries D95} (2017) 094503},
  [\href{https://arxiv.org/abs/1611.08285}{{\ttfamily 1611.08285}}].

\bibitem{Nonaka:2004pg}
C.~Nonaka and M.~Asakawa, \emph{{Hydrodynamical evolution near the QCD critical
  end point}}, \href{https://doi.org/10.1103/PhysRevC.71.044904}{\emph{Phys.
  Rev.} {\bfseries C71} (2005) 044904},
  [\href{https://arxiv.org/abs/nucl-th/0410078}{{\ttfamily nucl-th/0410078}}].

\bibitem{Guida:1996ep}
R.~Guida and J.~Zinn-Justin, \emph{{3-D Ising model: The Scaling equation of
  state}}, \href{https://doi.org/10.1016/S0550-3213(96)00704-3}{\emph{Nucl.
  Phys.} {\bfseries B489} (1997) 626--652},
  [\href{https://arxiv.org/abs/hep-th/9610223}{{\ttfamily hep-th/9610223}}].

\bibitem{Schofield:1969zz}
P.~Schofield, J.~D. Litster and J.~T. Ho, \emph{{Correlation Between Critical
  Coefficients and Critical Exponents}},
  \href{https://doi.org/10.1103/PhysRevLett.23.1098}{\emph{Phys. Rev. Lett.}
  {\bfseries 23} (1969) 1098--1102}.

\bibitem{Bluhm:2006av}
M.~Bluhm and B.~Kampfer, \emph{{Quasi-particle perspective on critical
  end-point}}, {\emph{PoS} {\bfseries CPOD2006} (2006) 004},
  [\href{https://arxiv.org/abs/hep-ph/0611083}{{\ttfamily hep-ph/0611083}}].

\bibitem{Cea:2015bxa}
L.~Cosmai, A.~Papa and P.~Cea, \emph{{Curvature of the pseudocritical line in
  (2+1)-flavor QCD with HISQ fermions}}, {\emph{PoS} {\bfseries LATTICE2015}
  (2016) 143}, [\href{https://arxiv.org/abs/1510.06847}{{\ttfamily
  1510.06847}}].

\bibitem{Bonati:2015bha}
C.~Bonati, M.~D'Elia, M.~Mariti, M.~Mesiti, F.~Negro and F.~Sanfilippo,
  \emph{{Curvature of the chiral pseudocritical line in QCD: Continuum
  extrapolated results}},
  \href{https://doi.org/10.1103/PhysRevD.92.054503}{\emph{Phys. Rev.}
  {\bfseries D92} (2015) 054503},
  [\href{https://arxiv.org/abs/1507.03571}{{\ttfamily 1507.03571}}].

\end{thebibliography}\endgroup

\end{document}